\documentclass[aps,prd,11pt,showpacs,superscriptaddress,amssymb,amsmath,nofootinbib]{revtex4-1}

\usepackage{hhline}
\usepackage{color}
\usepackage{epsfig}
\usepackage{graphicx}
\usepackage{float}%
\usepackage{epstopdf}
\usepackage{hyperref}
\usepackage[usenames,dvipsnames]{xcolor}
\usepackage{multirow,booktabs}
\usepackage[normalem]{ulem}
\usepackage[d]{esvect}
\usepackage[misc,geometry]{ifsym}
\usepackage{harpoon}
\usepackage{subcaption}
\usepackage{graphicx}
\usepackage{cleveref}

\begin{document}

	\title{Effects of gravitational lensing on neutrino oscillation in Hu-Sawicki f(R) gravity }
	\author{Ya-Ru Wang}
	\author{Ze-Wen Li}
	\author{Shu-Jun Rong}\email{rongshj@glut.edu.cn}
	\affiliation{College of Physics and Electronic Information Engineering, Guilin University of Technology, Guilin, Guangxi 541004, China}

	\begin{abstract}
		Gravitational lensing serves as a powerful probe of compact astrophysical objects and dark matter distributions. As relativistic counterparts to photons, neutrinos experiencing lensing offer a complementary means to investigate the properties of curved spacetimes. This paper studies neutrino oscillations within the spacetime geometry described by the Hu-Sawicki f(R) gravity model, focusing on the modifications induced by gravitational lensing. We calculate the oscillation phases for both radial and non-radial neutrino propagation and derive the corresponding flavor transition probabilities for 2-flavor and 3-flavor scenarios under the weak-field approximation. Our analysis demonstrates that the lensing-affected oscillation probabilities exhibit a clear dependence on the Hu-Sawicki model parameter $\lambda$ , the neutrino mass hierarchy, and the absolute value of the lightest neutrino mass.  Furthermore, extending the analysis beyond the weak-field regime reveals that strong-field gravitational lensing amplifies these effects. These results, while theoretical, indicate that future high-precision measurements of lensed neutrinos from compact astrophysical objects could, in principle, help test modified gravity models and constrain neutrino parameters, provided that experimental and wave-packet decoherence challenges are overcome.
		
	\end{abstract}
	

	\maketitle

	\section{Introduction}
	Neutrino oscillations  provide compelling hints for physics beyond the Standard Model through the phenomena of neutrino mass and flavor mixing \cite{Fukuda:1998mi, Ahmad:2002jz}. While oscillation probabilities depend primarily on mass-squared differences \cite{Pontecorvo:1957cp}, the absolute neutrino mass scale, its origin, and the mass ordering remain fundamental open questions \cite{deGouvea:2013on, Esteban:2020cvm}. These are not only essential for completing the picture of particle physics but also have profound implications for cosmology and astrophysics \cite{Lesgourgues:2012uu, Abazajian:2011dt}.
	
	Gravitational lensing, a key prediction of general relativity, has matured into a powerful astrophysical probe \cite{Bartelmann:2010, Schneider:2006}. It enables the study of dark matter distributions \cite{Massey:2010nd}, the detection of distant celestial objects, and tests of gravity in strong-field regimes \cite{Perlick:2004tq, Virbhadra:2000}. For neutrinos, especially at high-energies, extragalactic neutrinos observed by detectors such as IceCube \cite{Aartsen:2013bka, IceCube:2018dnn} and KM3NeT \cite{Adrian-Martinez:2016fdl}—propagation through the curved spacetime near compact objects may imprint lensing-induced signatures onto their flavor evolution \cite{Crocker:2004nh, Fornengo:1997}. Unlike electromagnetic radiation, neutrinos interact only weakly, traveling largely unimpeded over cosmological distances, thus serving as important messengers of both particle properties and spacetime geometry \cite{Ahlers:2010, Anchordoqui:2014}.
	
	Gravity plays a role in the propagation of neutrinos\cite{Wudka:1991tg,Ahluwalia:1996ev,Cardall:1996cd,Fornengo:1996ef,Piriz:1996mu,Crocker:2003cw,Zhang:2003pn,Lambiase:2005gt,Ren:2010yf,Visinelli:2014xsa,Lin:2019lko,Boshkayev:2020igc,Mandal:2021dxk,Capolupo:2020wlx,Swami:2020qdi,
	Buoninfante:2019der,Koutsoumbas:2019fkn,Pantig:2022gih,Swami:2022xet,Chakrabarty:2023kld,Alloqulov:2024sns}, particularly for those travelling around compact astrophysical objects. Modified theories of gravity, such as f(R) gravity model \cite{Sotiriou:2008rp, DeFelice:2010aj, Capozziello:2011et}, offer compelling alternatives to general relativity, particularly in addressing cosmological puzzles like late-time acceleration. The Hu-Sawicki f(R) gravity model \cite{Hu:2007nk}, notable for its ability to mimic a cosmological constant while introducing scalable curvature corrections, provides a rich framework for exploring strong-gravity phenomenology. Neutrino flavor transitions in such a spacetime may shed light on both neutrino intrinsic properties and deviations from general relativity \cite{Chakrabarty:2023, Swami:2020}.
	
In this context, following the methodology  proposed in Refs.\cite{Fornengo:1996ef,Swami:2020qdi},we systematically investigate how gravitational lensing in the spacetime of a the Hu-Sawicki f(R) gravity model affects neutrino oscillations. The choice of the Hu-Sawicki $f(R)$ model is motivated by its ability to mimic a cosmological constant while passing solar system tests, making it a well-studied alternative to general relativity.Going beyond previous studies limited to weak-field approximations , we derive the neutrino phase in fully covariant form for both radial and lensed non-radial trajectories. We compute oscillation probabilities for two and three-flavor systems, analyze their dependence on the model parameter $\lambda$
, the neutrino mass ordering, and the lightest neutrino mass, and compare weak-field and strong-field regimes. 

The paper is organized as follows. In \cref{sec:phase} we outline the theoretical framework of neutrino oscillations  in flat and curved spacetimes. Considering lensing effects by Hu-Sawicki f(R) gravity model on neutrinos, we derive the oscillation phases for radial and non-radial propagations. In \cref{sec:prob} the neutrino oscillation probabilities in 2-flavor and 3-flavor case are calculated under the weak-filed approximation, showing the impacts of neutrino masses and the Hu-Sawicki f(R) gravity parameter on flavor transitions. In \cref{sec:strong} We investigate the effect of strong-field gravitational lensing on neutrino oscillations including 2-flavor and 3-flavor. A conclusion is given in \cref{sec:conclusion}. Through out the paper, we take the unit $G=\hbar=c=1$.

	\section{Phases in neutrino oscillations}\label{sec:phase}
	\subsection{Phases in flat spacetime }
	We first outline the theory of neutrino oscillations in flat spacetime based on quantum mechanics.
	Neutrinos are produced and detected in flavor states $\left| \nu_{\alpha} \right\rangle$ through weak interactions, in which \(\alpha = e,\mu,\tau\) . The states are considered to be superpositions of mass eigenstates represented by $|\nu_i\rangle$, i.e.,
	\begin{equation}
		|\nu_{\alpha}\rangle = \sum_{i = 1}^{3}U_{\alpha i}^{\ast}|\nu_{i}\rangle,
		\label{eq:nu_i_to_alpha}
	\end{equation}
	where \(U_{\alpha i}\) is the $3\times 3$  Pontecorvo-Maki-Nakagawa-Sakata (PMNS) leptonic mixing matrix\cite{Pontecorvo:1957qd,Maki:1962mu,Pontecorvo:1967fh}.
	Supposing a neutrino propagating from the source $S$ at \(\left( t_{S},x_{S} \right)\ \) to the detector $D$ at \(\left( t_{D},x_{D} \right)\), the final mass state is given by
	\begin{equation}
		\left| \nu_{i}\left( t_{D},x_{D} \right)\rangle = \exp\left( - i\Phi_{i} \right) \right|\nu_{i}\left( t_{S},x_{S} \right)\rangle,
		\label{eq:plane_wave}
	\end{equation}
	where \(\Phi_{i}\) is the phase generated by the propagation. The probability of flavor oscillation is defined as follows
	\begin{equation}
		\label{eq:prob_alpha_to_beta}
		P_{\alpha\beta} = \mid \langle\nu_{\beta}|\nu_{\alpha}\left( t_{D},x_{D} \right)\rangle \mid^{2}= \sum_{i,j}^{}U_{\beta i}U_{\beta j}^{\ast}U_{\alpha j}U_{\beta i}^{\ast}\exp\left[-i\left( \Phi_{i} - \Phi_{j} \right) \right]. \\
	\end{equation}
	In flat spacetime, taking the plane-wave ansatz in 1-dimension for neutrinos, \({\ \Phi}_{i}\) is expressed by \cite{Akhmedov:2009rb,Akhmedov:2010ua}
	\begin{equation}
		\label{eq:flat_phase}
		\Phi_{i} = E_{i}\left( t_{D} - t_{S} \right) - p_{i}\left( x_{D} - x_{S} \right).
	\end{equation}
	The phase differences determining the oscillation probability are dependent on the squared mass difference of neutrinos \(\ \Delta m_{ij}^{2} = m_{i}^{2} - m_{j}^{2}\) as follows
	\begin{equation}
		\label{eq:flat_phase_diff}
		\Delta\Phi_{ij} \equiv \Phi_{i} - \Phi_{j} \simeq \frac{\Delta m_{ij}^{2}}{2E_{0}}\left|x_{D} - x_{S} \right|,
	\end{equation}
	where \(E_{0}\) is the average energy of relativistic neutrinos.

	\subsection{Phases in curved spacetime}
	In curved spacetime, the expression of the phase \(\Phi_{k}\) of neutrinos can be written in a covariant form \cite{Stodolsky:1978ks}
	\begin{equation}
		\label{eq:curved_phase}
		\Phi_{k} = \int_{S}^{D}p_{\mu}^{\left( k \right)}dx^{\mu},
	\end{equation}
	where
	\begin{equation}
		\label{eq:p_mu}
		{\ p}_{\mu}^{\left( k \right)} = m_{k}g_{\mu\nu}\frac{dx^{\nu}}{ds}
	\end{equation}
	is the canonical conjugate momentum to the coordinates \(x^{\mu}\), \(g_{\mu\nu}\) and $ds$ are the metric tensor and the line element of the curved spacetime, respectively. $(k)$ represents the $k$-th mass eigenstate. $m_k$ is the mass which satisfies the mass-shell condition
	\begin{equation}
		\label{eq:mass_shell}
		m_k^2=g^{\mu\nu}p_\mu^{(k)}p_\nu^{(k)}.
	\end{equation}
	
The metric we consider is obtained from the Hu-Sawicki $f(R)$ gravity model under the weak-field approximation. Assuming a static and spherically symmetric spacetime, the line element takes the following form \cite{Ellis:1973yv,Morris:1988cz}:
\begin{equation}
ds^2 = - A(r) dt^2 + B(r) dr^2 + r^2(d\theta^2 + \sin^2\!\theta\, d\phi^2),
\label{eq6}
\end{equation}
	with \cite{Mohan:2024hbr}
	\begin{align}
		A(r) = \frac{1}{B(r)} & = 1 - \frac{2M}{r} + \frac{m^2}{12} \left(\frac{n-2}{2c_2}\right)^{\!1/n}\!\!\! r^2 \nonumber\\[5pt]
		& = 1-\frac{2 M}{r} + \lambda\, r^2,
		\label{eq7}
	\end{align}
	\footnote{In the Hu-Sawicki model, $f(R)=R - m^2\frac{c_1 (R/m^2)^n}{c_2 (R/m^2)^n+1}$. The effective cosmological constant is $\Lambda = \frac{c_1}{2c_2}m^2$, and the parameter $\lambda$ in Eq.~(10) is given by $\lambda = \frac{m^2}{12}\left(\frac{n-2}{2c_2}\right)^{1/n}$. For details see \cite{Mohan:2024hbr}.}
	where $M$ is the balck hole mass parameter and
	$\lambda = m^2/12\left((n-2)/2c_2\right)^{\!1/n}$. Note that $\lambda=0$ reduces to the case of Schwarzschild spacetime.
	To calculate the phase Eq.~(\ref{eq:curved_phase}) we need to evaluate the canonical momenta $p_\mu^{(k)}$. Through out this paper, we restrict our discussions to neutrinos traveling in the $\theta=\pi/2$ equatorial plane, therefore $p_\theta^{(k)}=0$. Since the metric components don't depend on $t$ and $\phi$, the corresponding momenta are constants along the trajectory of neutrinos.  The nontrivial momenta and differentials are listed as follows
	\begin{equation}
		\label{eq:momenta}
		\begin{aligned}
			p_t^{(k)}&=m_k g_{tt}{dt\over ds} \equiv-E_k,   \qquad &{dt\over ds}&=-{E_k\over m_k g_{tt}},   \\
			p_r^{(k)}&=m_k g_{rr} {dr\over ds}\equiv p_k, \qquad &{ds\over dr}&={m_k g_{rr}\over p_k},   \\
			p_\varphi^{(k)}&= m_kg_{\varphi\varphi}{d\varphi\over ds}\equiv J_k,\qquad &{d\varphi\over ds}&={J_k\over m_k g_{\varphi\varphi}}.
		\end{aligned}
	\end{equation}
	
	\subsubsection{Radial propagation}
	If neutrinos propagate radially, the angle $\varphi$ stays constant $d\varphi=0$, therefore $J_k=0$. Using Eq.~(\ref{eq:curved_phase}), the phase in this scenario is \cite{Fornengo:1996ef}
	\begin{equation}
		\label{eq:curved_phi}
		\Phi_{k} = \int_{S}^{D}\left\lbrack - E_{k}\left( \frac{dt}{d r} \right)_{0} + p_{k}\right\rbrack dr,
	\end{equation}
	where $S$ and $D$ denote the source and detector of the neutrinos,
	respectively. The light-ray differential is written as
	\begin{equation}
		\label{eq:dt_dr_0}
		\left( \frac{dt}{dr} \right)_{0} = \frac{E_{0}}{p_{0}}\frac{B}{A},
	\end{equation}
	where \(E_{0}\) and \(p_{0}\) are the energy and momentum of a massless particle at infinity. The momentum $p_k$ obtained from the mass-shell condition Eq.~(\ref{eq:mass_shell}) reads
	\begin{equation}
		\label{eq:radial_pk}
		p_{k}= \pm \sqrt{\frac{B E_{k}^{2}}{A} - B m_{k}^{2}}
	\end{equation}
	In the massless case, we have $p_0=\pm E_0\sqrt{{B}/{A}}$. Substituting the above expressions into Eq.~(\ref{eq:curved_phi}), the phase is obtained as follow
	\begin{equation}
		\label{eq:radial_phase_int}
	\Phi_{k} = \pm \int_{S}^{D}E_{k}\sqrt{\frac{B}{A}}\left[-1 + \sqrt{1 - \frac{m_{k}^{2}A}{E_{k}^{2}}}\right]dr.
	\end{equation}
	Utilizing the expressions ${A,B}$ and the relativistic approximation\cite{Fornengo:1996ef}
	\begin{equation}
		\label{eq:relat_approx}
		E_k\simeq E_0 + \mathcal{O}\left({m_k^2\over 2E_0}\right),
	\end{equation}
	the phase is simplified as follow
	\begin{equation}
		\label{eq:radial_phase_expand_root}
		\begin{aligned}
			\Phi_{k} & = \pm E_{k}\left( - 1 + \sqrt{1 - \frac{m_{k}^{2}}{E_{k}^{2}}} \right)\left( r_D - r_S \right) \\
			& \approx \pm \frac{m_{k}^{2}}{2E_{0}}\left( r_D - r_S \right). \\
		\end{aligned}
	\end{equation}
	Note that the result derived from \cite{Fornengo:1996ef} is the same as the phase from the Schwarzschild black hole, while it is different from the phase shown in \cite{Godunov:2009ce}. In this paper, we follow the methodology proposed in \cite{Fornengo:1996ef,Swami:2020qdi}.
	
	\subsubsection{Non--radial propagation}
	If the gravitational lensing effect sets in, the neutrinos propagate non-radially. The angle $\varphi$ in this case is no longer constant along the trajectory, therefore the phase is also influenced by $J_k$. The phase in Eq.~(\ref{eq:curved_phase}) now is expressed as
	\begin{equation}
		\label{eq:non_rad_phase_ori}
		\Phi_k=\int_S^D\left[-E_k\left({dt\over dr}\right)_0+p_k+J_k\left({d\varphi\over dr}\right)_0\right],
	\end{equation}
	where the light-ray differentials are
	\begin{equation}
		\label{eq:non_rad_lr_diff}
		\left({dt\over dr}\right)_0={E_0\over p_0}{B\over A},\qquad \left({d\varphi
			\over dr}\right)_0={J_0\over p_0}{B\over D},
	\end{equation}
with ${D}=r^2 \sin^2\theta$.
	The angualr momentum $J_k$ is expressed as \cite{Weinberg:1972kfs,Fornengo:1996ef}
	\begin{equation}
		\label{eq:Jk_ori}
		J_k=E_k b v_k,
	\end{equation}
	where $b$ is the impact parameter and $v_k$ is the velocity of neutrino in the $k$-th mass eigenstate.We define that where the metric is flat, the velocity can be written as
	\begin{equation}
		\label{eq:v}
		v_k={\sqrt{E_k^2-m_k^2}\over E_k}\simeq 1-{m_k^2\over 2E_k^2},
	\end{equation}
	and the angular momentum can be approximated as
	\begin{equation}
		\label{eq:jk}
		J_k\simeq E_k b(1-{m_k^2\over 2E_k^2})=b(E_k-{m_k^2\over 2E_k}),\quad J_0=E_0 b,
	\end{equation}
	where $J_0$ is the angualr momentum of the massless particle. To calculate $p_k, p_0$, adopting again the mass-shell relation Eq.~(\ref{eq:mass_shell}) with non-vanishing $J_k$, we get
	\begin{equation}
		\label{eq:pk}
		p_k=\pm E_k\sqrt{{1\over{A}}-{b^2\over {D}}-\left(1-{b^2\over{D}}\right){m_k^2\over E_k^2}},\quad p_0=\pm E_0\sqrt{{1\over {A}}-{b^2\over {D}}}.
	\end{equation}
	Substituting the above results into Eq.~(\ref{eq:non_rad_phase_ori}), the phase is reduced to
	\begin{equation}
		\label{eq:phase_non_rad_in}
		\Phi_{k} = \pm \frac{m_{k}^{2}}{2E_{0}}\int_{S}^{D}\sqrt{{AB}}\left(1 -\frac{b^{2}{A}}{{D}} \right)^{- \frac{1}{2}}dr,
\end{equation}
	where the relativistic approximation is used.
If we define the point of closest approach to occur at  $r=r_0$, the corresponding phase may be written in the form
\begin{equation}\label{eq13}
	\Phi_{k}\left(r_{S} \rightarrow r_{0} \rightarrow r_{D}\right) = \frac{m_{k}^{2}}{2E_{0}} \int_{r_{0}}^{r_{S}} \sqrt{\frac{AB}{1 - \frac{b^{2}A}{r^{2}}}} dr + \frac{m_{k}^{2}}{2E_{0}} \int_{r_{0}}^{r_{D}} \sqrt{\frac{AB}{1 - \frac{b^{2}A}{r^{2}}}} dr.
	\end{equation}
Considering neutrino propagation at the shortest radius, we can conclude
	\begin{equation}\label{eq:noo}
		\left(\frac{dr}{d\phi}\right)_{0} = \frac{p_0(r_0)D}{J_0 B} = 0.
	\end{equation}
	Substituting Eq.~(\ref{eq:jk}) and Eq.~(\ref{eq:pk})into Eq.~(\ref{eq:noo}), we can obtain
	\begin{equation}\label{eq:15}
		\frac{\sqrt{ \frac{B}{\mathcal{A}} - \frac{B b^2}D } \times D }{b \times B}=0.
	\end{equation}
This equation is used to determine the relationship between $b$ and  $r_0$.
	\section{Neutrino oscillation probabilities in the weak-field regime}\label{sec:prob}
	\subsection{Theoretical results}
	By substituting Eq.~(\ref{eq7}) into Eq.~(\ref{eq13}), we obtain the phase of Hu-Sawicki $f(R)$ gravity model in the form
		\begin{equation}
		\label{eq:phisd}
		\begin{split}
\Phi_k = \frac{m_k^2}{2E_0} \int_{r_0}^{r_S} \frac{dr}{\sqrt{1 - \frac{b^2}{r^2} + \frac{2M b^2}{r^3} - \lambda b^2}} + \frac{m_k^2}{2E_0} \int_{r_0}^{r_D} \frac{dr}{\sqrt{1 - \frac{b^2}{r^2} + \frac{2M b^2}{r^3} - \lambda b^2}}.
	\end{split}
\end{equation}
Employing the weak-field approximation $M/r\ll1$ and expanding to the second order, the phase can be further expressed as
\begin{equation}\label{eq18}
	\begin{split}
		\Phi_{k}\left(r_{S} \rightarrow r_{0} \rightarrow r_{D}\right) = \frac{m_{k}^{2}}{2E_{0}} \int_{r_{0}}^{r_{S}} \left( \frac{1}{\sqrt{\frac{-b^2 + r^2 - b^2 r^2 \lambda}{r^2}}} + \frac{b^2 M}{r \sqrt{\frac{-b^2 + r^2 - b^2 r^2 \lambda}{r^2}} \left(b^2 - r^2 + b^2 r^2 \lambda\right)}\right) dr\\
		+\frac{m_{k}^{2}}{2E_{0}} \int_{r_{0}}^{r_{D}} \left( \frac{1}{\sqrt{\frac{-b^2 + r^2 - b^2 r^2 \lambda}{r^2}}} + \frac{b^2 M}{r \sqrt{\frac{-b^2 + r^2 - b^2 r^2 \lambda}{r^2}} \left(b^2 - r^2 + b^2 r^2 \lambda\right)}\right) dr
	\end{split}
\end{equation}
By integrating Eq.~(\ref{eq18}), it follows that
\begin{equation}
	\begin{aligned}
	\label{eq:phizuihzhong}
	\Phi_k = \frac{m_k^2}{2E_0} \left. \left[ \frac{-r(M + r) + b^2(1 + Mr\lambda + r^2\lambda)}{r\sqrt{1 + b^2\left(-\frac{1}{r^2} - \lambda\right)} \left(-1 + b^2\lambda\right)} \right] \right|_{r_0}^{r_S}+\frac{m_k^2}{2E_0} \left. \left[ \frac{-r(M + r) + b^2(1 + Mr\lambda + r^2\lambda)}{r\sqrt{1 + b^2\left(-\frac{1}{r^2} - \lambda\right)} \left(-1 + b^2\lambda\right)} \right] \right|_{r_0}^{r_D}.
	\end{aligned}
\end{equation}
    We now turn to the relationship between $b$ and $r_0$. By combining Eq.~(\ref{eq7}) with Eq.~(\ref{eq:15}), we obtain
    \begin{equation}
    	\begin{aligned}
    	\label{eq31}
   	b &= \pm \frac{r_0}{\sqrt{1+\lambda r_0^2 - \frac{2M}{r_0}}} \\
   &= \pm \frac{r_0}{\sqrt{1+\lambda r_0^2\left(1 - \frac{2M}{r_0(1+\lambda r_0^2)}\right)}}
\end{aligned}
\end{equation}
In the weak-field approximation, the expression reduces to
 \begin{equation}
	\begin{aligned}
b &= \frac{r_0}{\sqrt{1+\lambda r_0^2}} \left(1+\frac{M}{r_0(1+\lambda r_0^2)}\right) \quad (b>0, r_0>0) \\
-b &= \frac{r_0}{\sqrt{1+\lambda r_0^2}} \left(1+\frac{M}{r_0(1+\lambda r_0^2)}\right) \quad (b<0, r_0>0).\\
	\end{aligned}
\end{equation}
	
	A neutrino flavor state $|\nu_\alpha\rangle$ produced at the source could change into the state $|\nu_\beta\rangle$ at the detector. It's oscillation probability is expressed as
	\begin{equation}
		\label{eq:EW_phase}
		\mathcal{P}_{\alpha\beta} = \left| \langle\nu_{\beta}^D \right|\nu_{\alpha}^S\rangle|^{2} = \left| N \right|^{2}\sum\limits_{i,j}U_{\beta i}U_{\beta j}^{\ast}U_{\alpha j}U_{\alpha i}^{\ast}\sum\limits_{p,q}\exp\left( - i\Delta\Phi_{ij}^{pq} \right),
	\end{equation}
	where $i,j$ represents the mass eigenstates and $p,q$ denote different paths neutrino can take.
	\begin{equation}
		\label{eq:norm}
		\left| N \right|^{2} = \left(\sum\limits_{i}\left|U_{\alpha i} \right|^{2}\sum\limits_{p,q}\exp\left( -i\Delta\Phi_{ii}^{pq} \right) \right)^{-1}\\
	\end{equation}
	is the normalization constant.
	\subsection{Two flavor neutrino}
\begin{figure}[hb]
		\centering
		\includegraphics[width=0.5\linewidth]{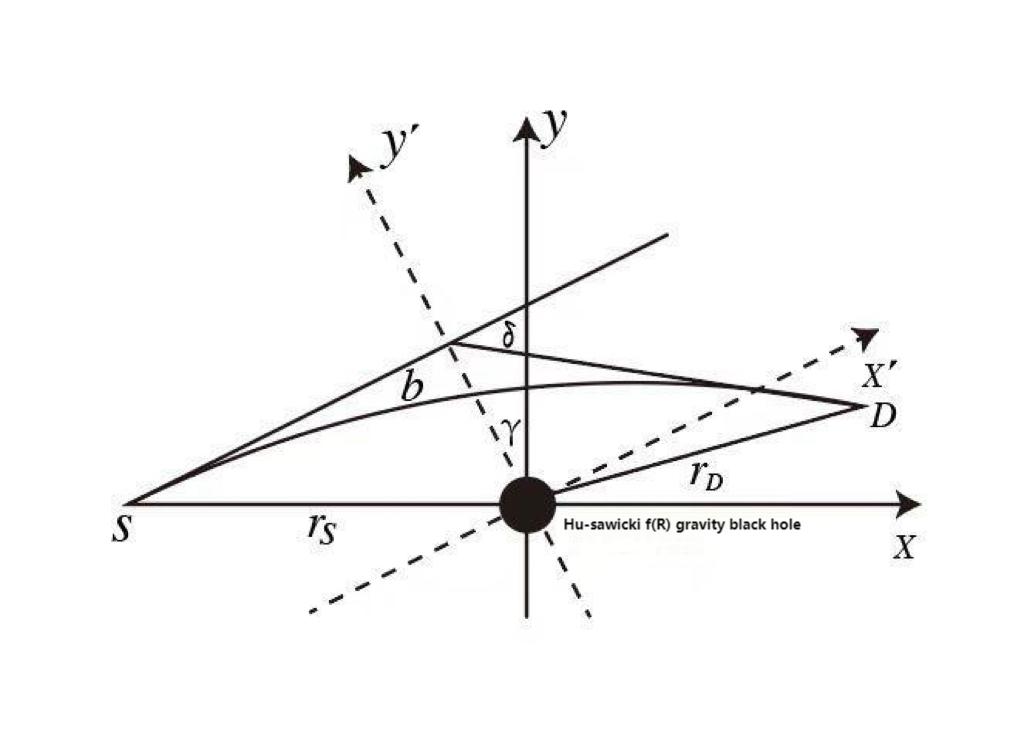}
		\caption{Illustration of gravitational lensing caused by an Hu-Sawicki $f(R)$ gravity model. $S$ is the source, $D$ is the detector. $b$ is the impact factor, $\delta$ is the deflection angle and $\gamma$ marks the misalignment of the coordinates $(x,y)$ and $(x',y')$.
			For the original construction of the plot, see \cite{Swami:2020qdi} and the adapted versions see \cite{Chakrabarty:2021bpr,Chakrabarty:2023kld,Alloqulov:2024sns}.}
		\label{fig:lens}
	\end{figure}

\begin{figure}
		\centering
		\begin{subfigure}{\textwidth}
			\includegraphics[width=16cm,height=6cm]{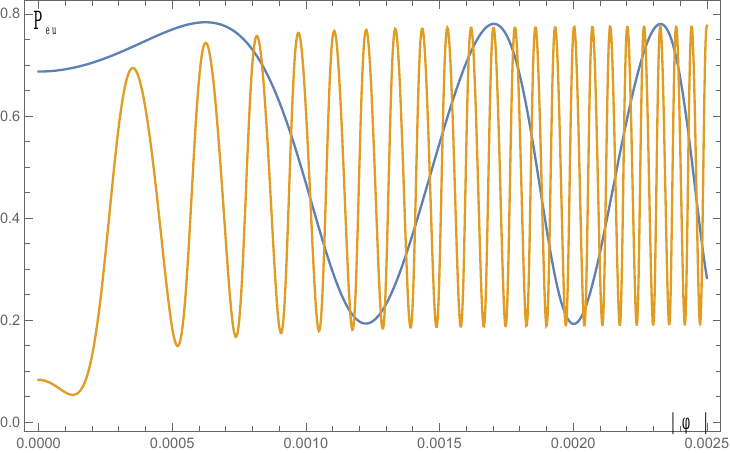}
			\caption{Normal ordering}
			\label{fig:2f_pi6_m2>0_m1=0}
		\end{subfigure}\hfill
		\begin{subfigure}{\textwidth}
			\includegraphics[width=16cm,height=6cm]{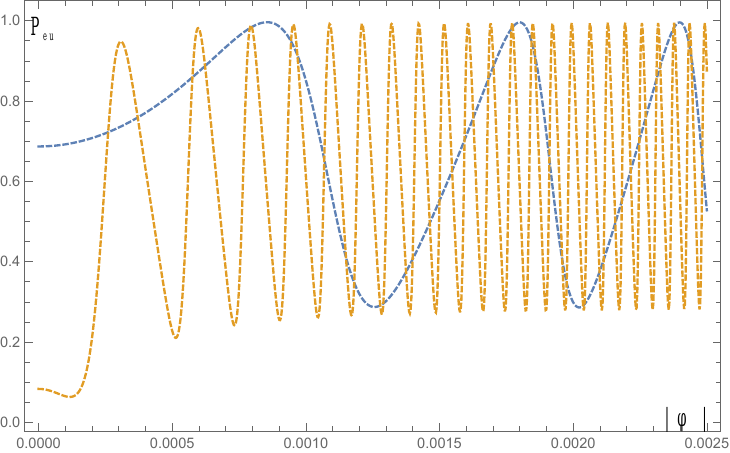}
			\caption{Inverted ordering}
			\label{fig:2f_pi6_m2<0_m1=0}
		\end{subfigure}\hfill
		\begin{subfigure}{\textwidth}
			\includegraphics[width=16cm,height=6cm]{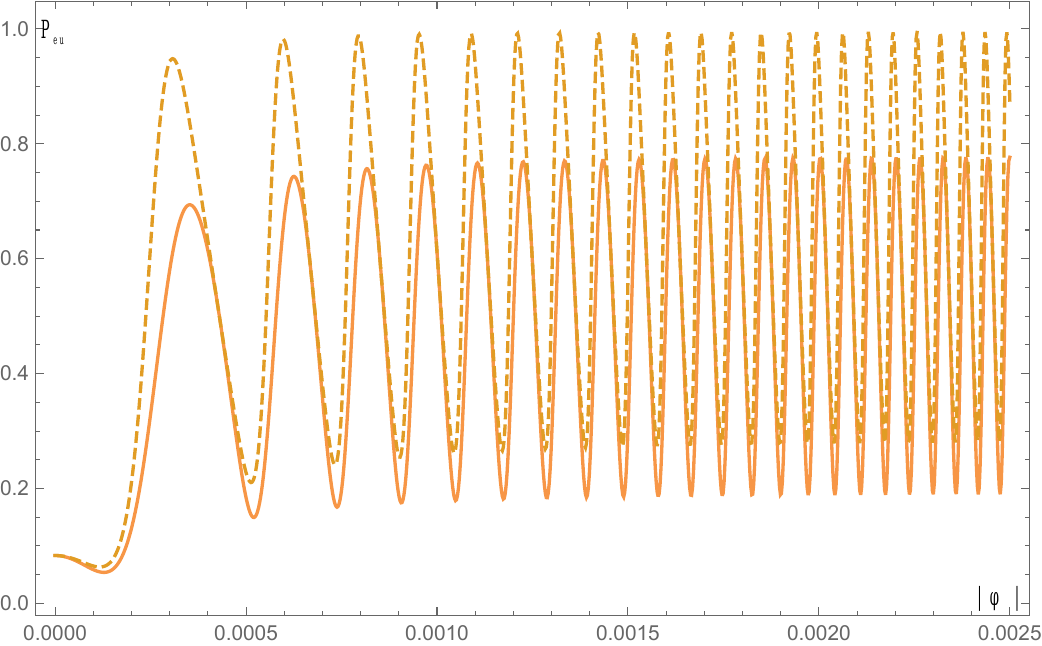}
			\caption{Normal ordering and Inverted ordering}
			\label{fig:2f_pi6_m2<0_m2=0}
		\end{subfigure}\hfill
		\caption{Oscillation probability of the two flavor case including the lensing effects of Hu-Sawicki $f(R)$ gravity model under weak-field. The neutrino mixing angle is chosen to be $\alpha=\pi/5$, lightest neutrino mass $m_{l}$ is 0, $r_D=10^8$ km, $r_S=10^5 r_D$, $E_0=10$ MeV, $|\Delta m^2|=10^{-3} \text{ eV}^2$.The blue curves correspond to $\lambda=0$, and the orange curves correspond to $\lambda=10^{-26}\,\text{m}^{-2}$. The solid curve represents the normal ordering, while the dashed curve represents the inverted ordering.}
		\label{fig:2f_pi6_1m}
	\end{figure}

	Before discussing the numerical results we need to calculate the impact factor. It is obtained from the relation between the deflection angle $\delta$ and the misalignment angle $\gamma$.  Let us define a Cartesian coordinate system $(x,y)$ with the origin at the black hole, see Fig.~\ref{fig:lens}.
The source and the detector are located at $(x_S,y_S)$ and $(x_D,y_D)$ in this system with radius $r_S,r_D$, respectively. Rotating the system with an angle $\gamma$, we arrive at $(x',y')$, which are related to the old ones through \(y^{\prime} = - x\sin\gamma + y\cos\gamma\) and \(x^{\prime} = x\cos\gamma + y\sin\gamma\). When $\gamma=0$, the source and the detector are collinear. The angel can be expressed in terms of the impact factor as $\sin\gamma=b/r_S$. In the rotated frame, the deflection angle $\delta$ is related to the impact parameter through
	\begin{equation}
		\label{eq:delta}
		\delta\sim-{y_D'-b\over x_D'}.
	\end{equation}
	Note that in the second step the deflection angle of neutrinos in the Hu-Sawicki $f(R)$ gravity model is as follows\cite{Mohan:2024hbr},
	\begin{equation}
		\label{eq:delta_EW}
	\delta = \frac{15\pi M^2}{4b^2} + \frac{15\pi \lambda M^2}{4} + \frac{4M}{b} - 3\lambda b M.
	\end{equation}
	Substituting $x_D',y_D'$ and $r_0$ into the above equation, and neglecting the term $ \frac{5\pi M^2}{2b^2}$, the impact parameter $b$ is solved through the polynomial equation
	\begin{equation}
		\label{eq:b_eq}
		-b - \frac{b x_D}{r_s} + \frac{b\left(\frac{4M}{b} - 3\lambda b M\right)x_D}{r_s} + \sqrt{1 - \frac{b^2}{r_s^2}}\left(\left(\frac{4M}{b} - 3\lambda b M\right)x_D + y_D\right) = 0.
	\end{equation}
	The impact parameters $b_1,b_2$ are the two real solutions to the above equation, which are in turn functions of the distance $r_{S,D}$ and the Hu-Sawicki $f(R)$ gravity model parameter $\lambda$.
	To illustrate the effect of gravitational lensing quantitatively, we consider the parameters of the Sun-Earth system. The detector is assumed to have a circular trajectory with $x_D=r_D \cos\phi$, $y_D=r_D \sin\phi$. The parameters are taken to be $r_D=10^8$ km, $r_S=10^5r_D$,$
	M = 1M_\odot$, $E_0=10$ MeV and we fix the squared mass difference to be $|\Delta m^2|=10^{-3}\text{ eV}^2$. We solve the impact parameters $b_1,b_2$ using eq.~(\ref{eq:b_eq}) numerically and calculate the probability with the angle $\phi$ in range $[0,0.0025]$.
	We take two values of $\lambda$ and plot the probabilites in fig.~\ref{fig:2f_pi6_1m}. The blue curves correspond to $\lambda=0$, and the orange curves correspond to $\lambda=10^{-26}\,\text{m}^{-2}$. The solid curve represents the normal ordering, while the dashed curve represents the inverted ordering. The same color and line-style conventions are adopted in all subsequent figures.

\begin{figure}
		\centering
		\begin{subfigure}{\textwidth}
			\includegraphics[width=16cm,height=6cm]{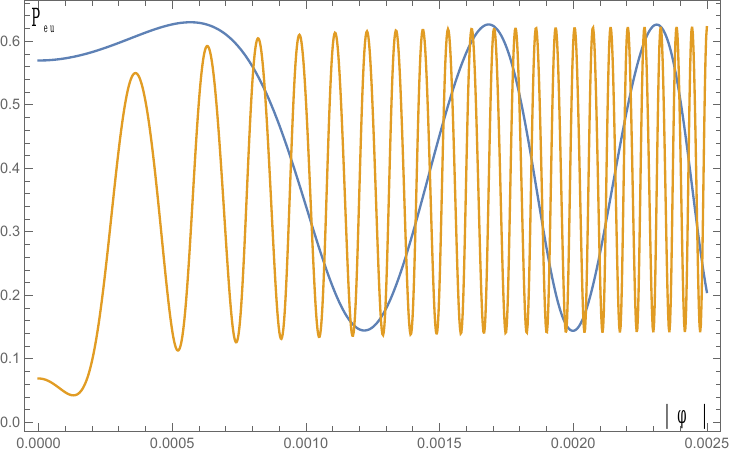}
			\caption{Normal ordering}
			\label{fig:2f_pi6_m2>0_m1=0}
		\end{subfigure}\hfill
		\begin{subfigure}{\textwidth}
			\includegraphics[width=16cm,height=6cm]{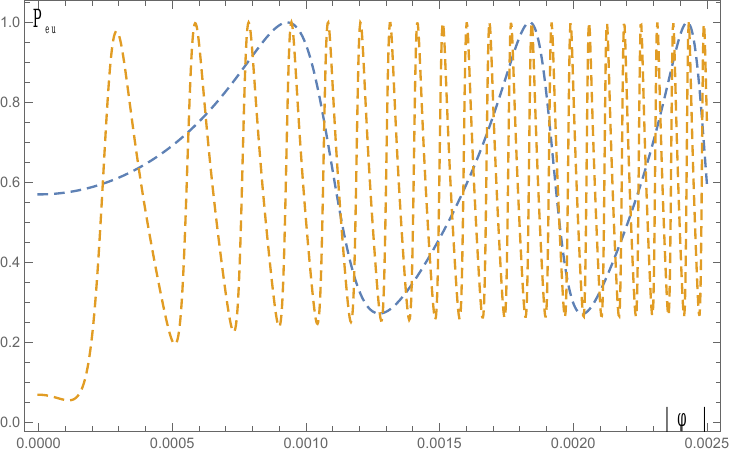}
			\caption{Inverted ordering}
			\label{fig:2f_pi6_m2<0_m1=0}
		\end{subfigure}\hfill
		\begin{subfigure}{\textwidth}
			\includegraphics[width=16cm,height=6cm]{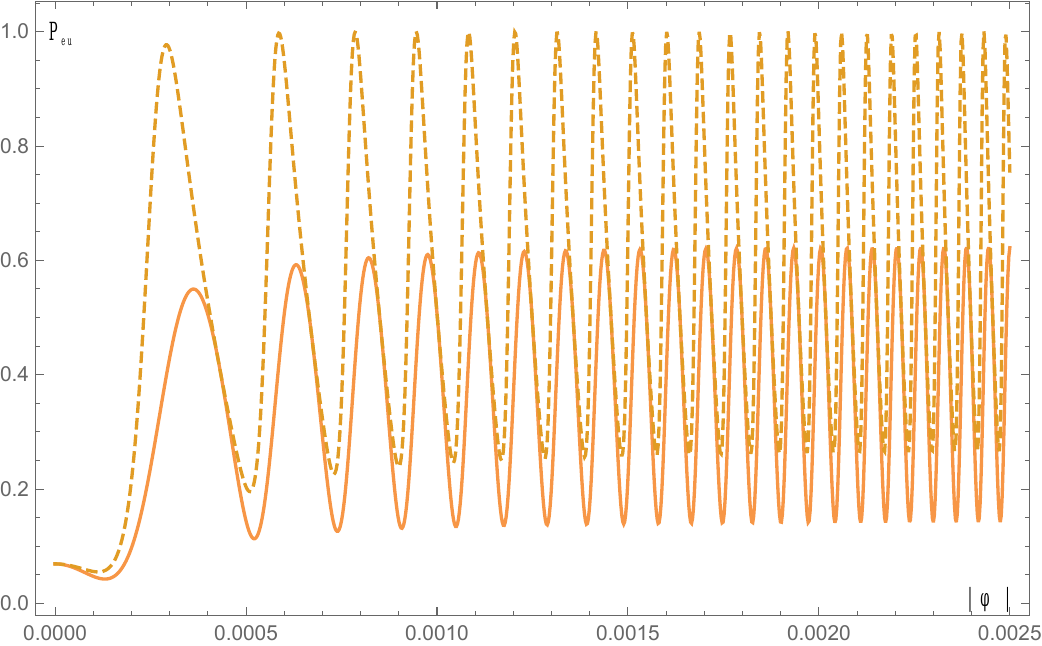}
			\caption{Normal ordering and Inverted ordering}
			\label{fig:2f_pi6_m2<0_m2=0}
		\end{subfigure}\hfill
		\caption{Oscillation probability of the two flavor case including the lensing effects of Hu-Sawicki $f(R)$ gravity model under weak-field . The neutrino mixing angle is chosen to be $\alpha=\pi/6$, lightest neutrino mass $m_{l}$ is 0, $r_D=10^8$ km, $r_S=10^5 r_D$, $E_0=10$ MeV, $|\Delta m^2|=10^{-3} \text{ eV}^2$.Parameter descriptions are the same as in fig.~\ref{fig:2f_pi6_1m}}
			\label{fig:2f_pi6_1m2}
	\end{figure}

	The oscillation probability is  sensitive to the mass hierarchy of  neutrinos. From fig.~\ref{fig:2f_pi6_1m}
	one can observe that the inverted ordering exhibits a larger amplitude compared to the normal ordering, and a clear difference is also evident between the cases with and without the parameter $\lambda$.We also calculated another case when the neutrino mixing angle is $\alpha=\pi/6$ in fig.~\ref{fig:2f_pi6_1m2}.
	Comparing to the $\alpha=\pi/6$ case, the amplitude in the $\alpha=\pi/5$ case is slightly higher in the case of normal mass ordering where the period stays almost the same.

\begin{figure}
		\centering
		\begin{subfigure}{\textwidth}
			\includegraphics[width=16cm,height=6cm]{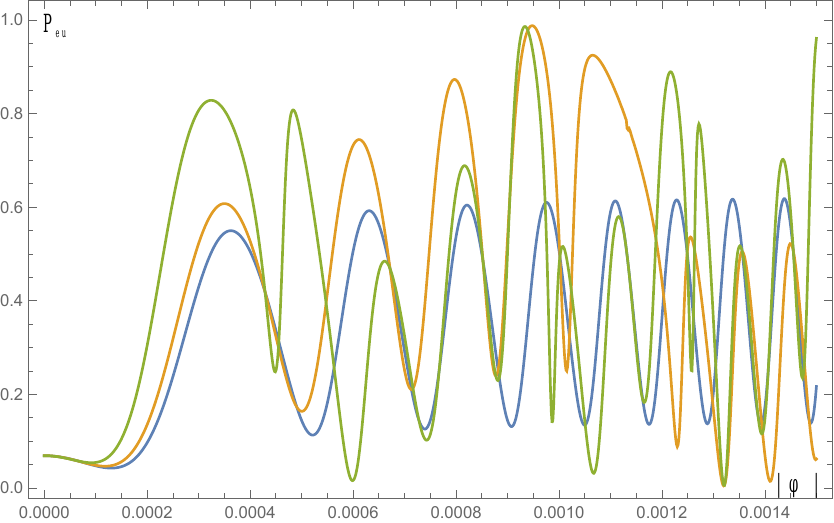}
			\caption{Normal ordering}
			\label{fig:2f_NO}
		\end{subfigure}
		\begin{subfigure}{\textwidth}
			\includegraphics[width=16cm,height=6cm]{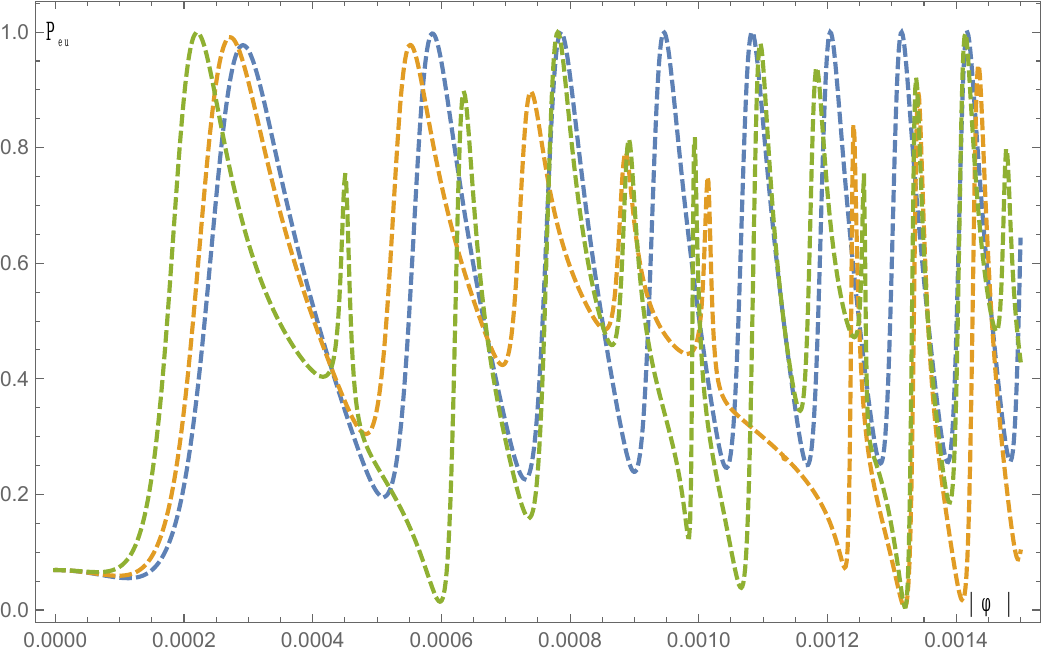}
			\caption{Inverted ordering}
			\label{fig:2f_IO}
		\end{subfigure}
		\caption{Oscillation probability of two flavor case $\nu_e\to\nu_\mu$. Top and bottom panel corresponds to normal ordering and inverted ordering, respectively. We vary in each plot the mass of the lowest neutrino, which we donote as $m_1$, from 0 eV(Blue line), 0.01 eV(yellow line) and 0.02eV(green line).The solid curve represents the normal ordering, while the dashed curve represents the inverted ordering. The neutrino mixing angle are chosen to be $\alpha=\pi/6$,  $r_D=10^8$ km, $r_S=10^5 r_D$, $E_0=10$ MeV, $|\Delta m^2|=10^{-3} \text{ eV}^2$.}
		\label{fig:2f_pi6_3m}
	\end{figure}

 The value of the lightest neutrino mass has significant impact on the oscillation probability. As shown in fig.~\ref{fig:2f_pi6_3m}, turning on the lowest mass may result in a shorter period. As the mass increases, the oscillation curve are distorted completely. Again we can see that the probability oscillates differently for different mass hierarchies.
	
	\subsection{Three flavor case}
	In this section we consider the case of three neutrino flavors and study the transitions $\nu_e\to\nu_\mu$, $\nu_e\to\nu_\tau$ and $\nu_\mu\to \nu_\tau$. The mixing matrix is the PMNS matrix, parametrized by three mixing angles $\theta_{12}$, $\theta_{13}$, $\theta_{23}$ and a phase $\delta_{CP}$. The mixing parameters are taken to be $\theta_{12}= 33.68^\circ$ $(\theta_{12} = 33.68^\circ)$, $\theta_{13}=8.56^\circ $ $(\theta_{13} = 8.59^\circ)$, $\theta_{23}= 43.3^\circ $ $(\theta_{23} = 47.9^\circ)$, $\delta_{\text{CP}} = 212^\circ$ $\left( \delta_{\text{CP}} = 274^\circ \right)$. The mass hierarchies are taken as $\Delta m_{21}^{2} = 7.49 \times 10^{- 5}\text{eV}^{2}$ $\left( \Delta m_{21}^{2} = 7.49 \times 10^{- 5}\text{eV}^{2} \right)$, $\Delta m_{31}=2.513 \times 10^{- 3}\text{eV}^{2}\left( \Delta m_{32}^{2} = -2.484 \times 10^{- 3}\text{eV}^{2} \right)$ for normal (inverted) ordering \cite{Esteban:2024eli}.

\begin{figure}
		\centering
		\begin{subfigure}{\textwidth}
			\includegraphics[width=16cm,height=6cm]{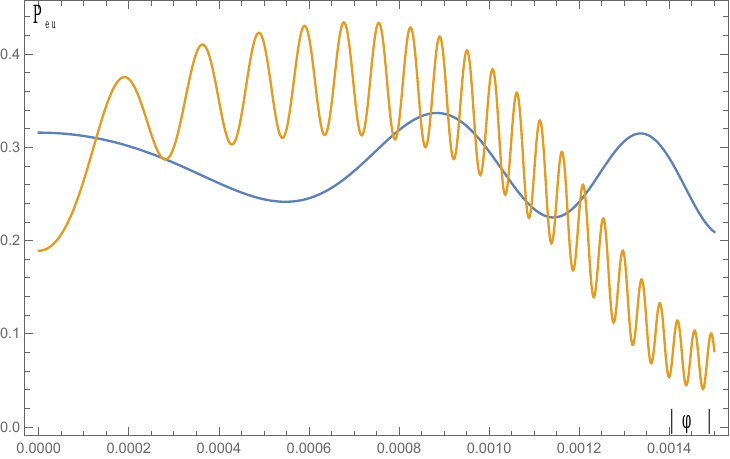}
			\caption{$\nu_e\to\nu_\mu$}
			\label{fig:3f_e_mu_NO}
		\end{subfigure}
		\hfill
		\begin{subfigure}{\textwidth}
			\includegraphics[width=16cm,height=6cm]{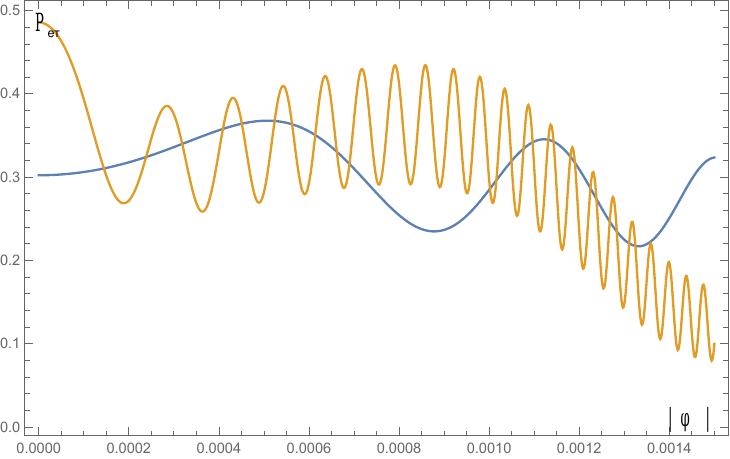}
			\caption{$\nu_e\to\nu_\tau$}
			\label{fig:3f_e_tau_NO}
		\end{subfigure}\hfill
		\begin{subfigure}{\textwidth}
			\includegraphics[width=16cm,height=6cm]{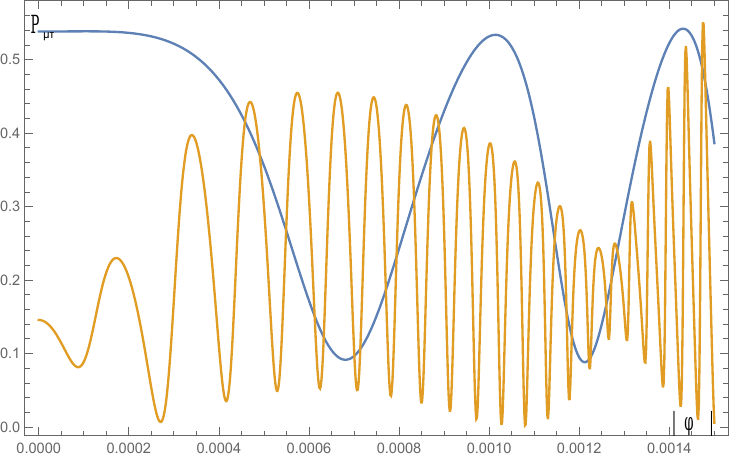}
			\caption{$\nu_\mu\to\nu_\tau$}
			\label{fig:3f_mu_tau_NO}
		\end{subfigure}
		\caption{Oscillation probability of the three flavor neutrino (normal ordering). From top to bottom correspond to $\nu_e\to\nu_\mu$, $\nu_e\to\nu_\tau$ and $\nu_\mu\to\nu_\tau$, respectively. $r_D=10^8$ km, $r_S=10^5 r_D$, $E_0=10$ MeV.The blue curves correspond to $\lambda=0$, and the orange curves correspond to $\lambda=10^{-26}\,\text{m}^{-2}$.}
		\label{fig:3f_NO}
	\end{figure}

\begin{figure}
		\centering
		\begin{subfigure}{\textwidth}
			\includegraphics[width=16cm,height=6cm]{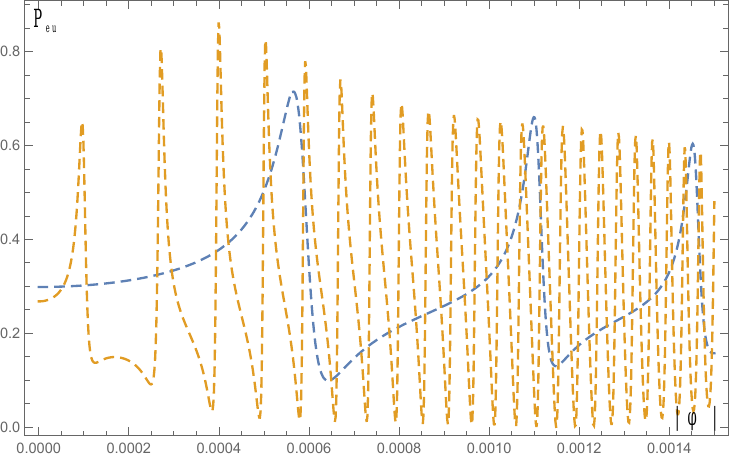}
			\caption{$\nu_e\to\nu_\mu$}
			\label{fig:3f_e_mu_IO}
		\end{subfigure}
		\hfill
		\begin{subfigure}{\textwidth}
			\includegraphics[width=16cm,height=6cm]{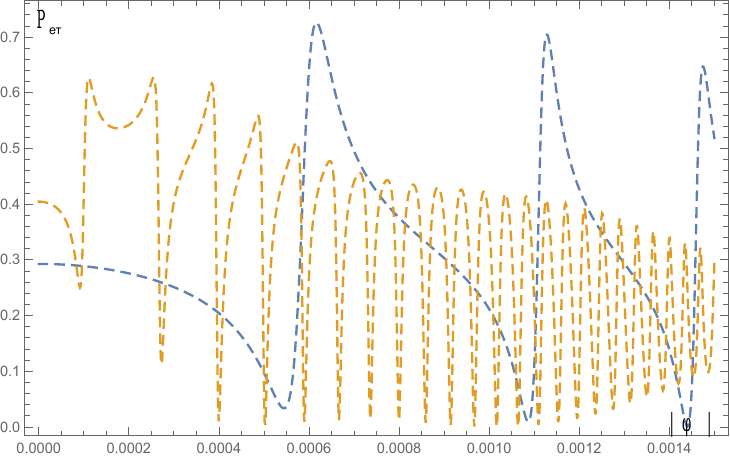}
			\caption{$\nu_e\to\nu_\tau$}
			\label{fig:3f_e_tau_IO}
		\end{subfigure}\hfill
		\begin{subfigure}{\textwidth}
			\includegraphics[width=16cm,height=6cm]{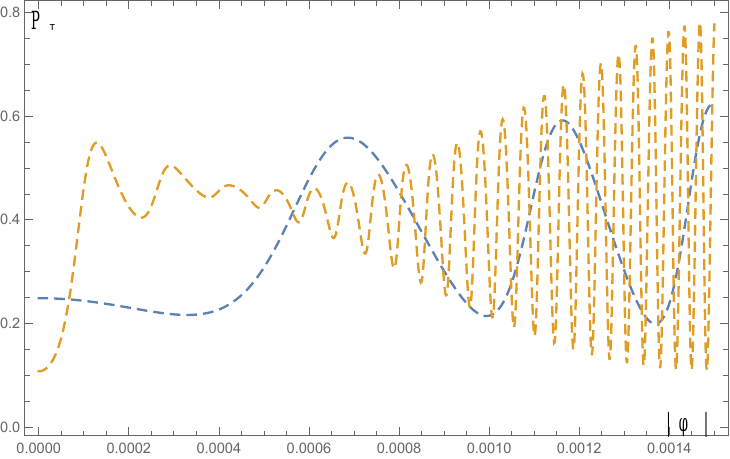}
			\caption{$\nu_\mu\to\nu_\tau$}
			\label{fig:3f_mu_tau_IO}
		\end{subfigure}
		\caption{Oscillation probability of the three flavor neutrino (inverted ordering). From top to bottom correspond to $\nu_e\to\nu_\mu$, $\nu_e\to\nu_\tau$ and $\nu_\mu\to\nu_\tau$, respectively. $r_D=10^8$ km, $r_S=10^5 r_D$, $E_0=10$ MeV.Parameter descriptions are the same as in fig.~\ref{fig:3f_NO}}
		\label{fig:3f_IO}
	\end{figure}

We consider again the Sun-Earth system where lightest neutrino's mass is 0, $r_D=10^8$ km, $r_S=10^5 r_D$,$M = 1M_\odot$, $E_0=10$ MeV and plot the oscillation probability against the angle $\phi\in[0,0.0015]$ for $\lambda=0,10^{-26}$ in fig.~\ref{fig:3f_NO} for normal mass ordering and fig.~\ref{fig:3f_IO} for inverted mass ordering. The process $\nu_\mu\to \nu_\tau$  shows a distinguished oscillation profile comparing to the others. We also see that the existence of Hu-Sawicki $f(R)$ gravity model changes the oscillation profile obviously comparing to the case of  Schwarzschild spacetime, especially for inverted ordering. This observation provides an opportunity  to examine  spacetime  property through neutrino oscillations and also a way to discriminate the mass hierarchy of  neutrinos using gravity.
	
	\section{Neutrino oscillation probabilities in the strong-field regime}\label{sec:prob2}
		\label{sec:strong}
\begin{figure}
		\centering
		\begin{subfigure}{\textwidth}
			\includegraphics[width=16cm,height=6cm]{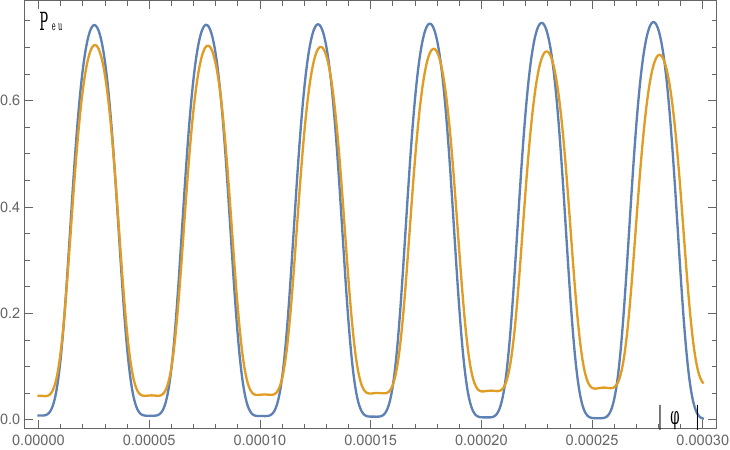}
			\caption{Normal ordering}
			\label{fig:9}
		\end{subfigure}\hfill
		\begin{subfigure}{\textwidth}
			\includegraphics[width=16cm,height=6cm]{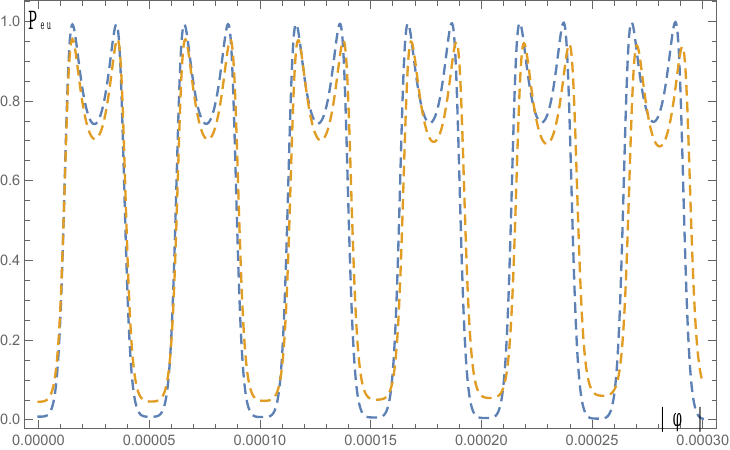}
			\caption{Inverted ordering}
			\label{fig:10}
		\end{subfigure}\hfill
		\caption{Oscillation probability of the two flavor case $\nu_e\to\nu_\mu$ including the lensing effects of Hu-Sawicki $f(R)$ gravity model under strong-field.The neutrino mixing angle are chosen to be $\alpha=\pi/5$,  $r_D=10^8$ km, $r_S=10^5 r_D$, $E_0=10$ MeV, $|\Delta m^2|=10^{-3} \text{ eV}^2$.The blue lines correspond to the zero $\lambda$ value, while the yellow lines represent the case with the parameter being $10^{-29}\text{m}^{-2}$and the solid curve represents the normal ordering, while the dashed curve represents the inverted ordering.}
		\label{fig:strong 2}
	\end{figure}

\begin{figure}
		\centering
		\begin{subfigure}{\textwidth}
			\includegraphics[width=16cm,height=6cm]{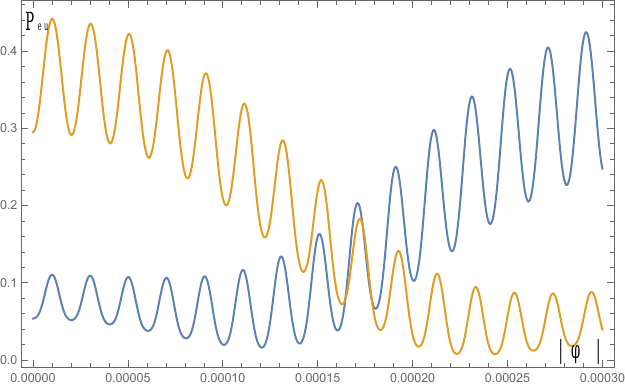}
			\caption{Normal ordering}
			\label{fig:17}
		\end{subfigure}
		\hfill
		\begin{subfigure}{\textwidth}
			\includegraphics[width=16cm,height=6cm]{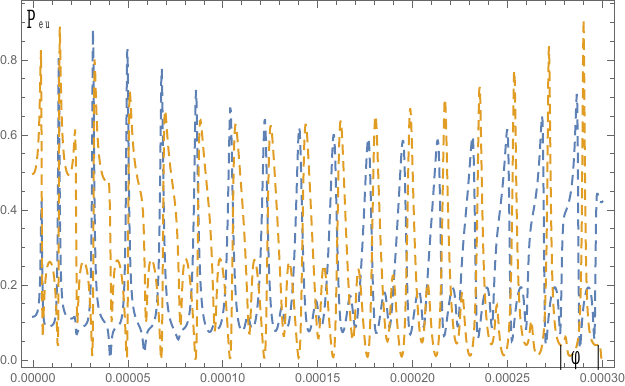}
			\caption{Inverted ordering}
			\label{fig:18}
		\end{subfigure}\hfill
			\begin{subfigure}{\textwidth}
			\includegraphics[width=16cm,height=6cm]{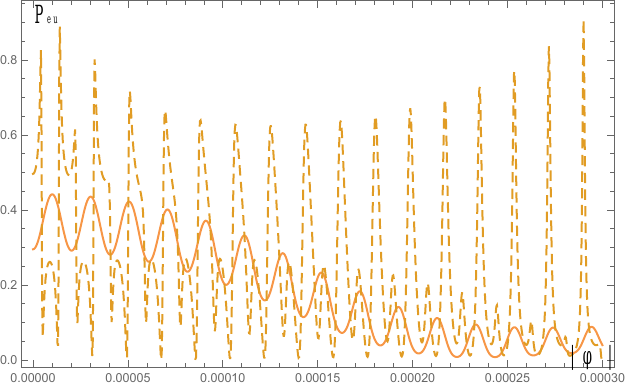}
			\caption{Normal ordering and Inverted ordering}
			\label{fig:19}
		\end{subfigure}
		\hfill
		\caption{Oscillation probability of the three flavor case $\nu_e\to\nu_\mu$ including the lensing effects of Hu-Sawicki $f(R)$ gravity model under strong-field. The neutrino mixing angle are chosen to be $r_D=10^8$ km, $r_S=10^5 r_D$, $E_0=10$ MeV.Parameter descriptions are the same as in fig.~\ref{fig:strong 2}}
		\label{fig:strong 3}
	\end{figure}

   In this section, we explore neutrino oscillations beyond the weak-field regime in the Hu-Sawicki $f(R)$ gravity model. We perform no approximations to the oscillation phase but to keep the integral as it is. Using the deflection angle in Eq.~(\ref{eq:delta_EW}) without neglecting the second term,
 we integrate directly the integral in Eq.~(\ref{eq:phase_non_rad_in}).	
We consider the distance parameters as in the previous sections. The detector is assumed to have a circular trajectory with $x_D=r_D \cos\phi$, $y_D=r_D \sin\phi$. The parameters are taken to be $r_D=10^8$ km, $r_S=10^5r_D$, $E_0=10$ MeV. We solve the impact parameters $b_1,b_2$ using Eq.~(\ref{eq:b_eq}) numerically and calculate the probability with the angle $\phi$ in range $[0,0.0005]$. The oscillation probabilities are shown for the two-flavor case in Fig.~\ref{fig:strong 2} and three-flavor case in Fig.~\ref{fig:strong 3}. The blue lines correspond to the zero $\lambda$ value, while the yellow lines represent the case with the parameter being $10^{-29}\text{m}^{-2}$and the solid curve represents the normal ordering, while the dashed curve represents the inverted ordering.

	In the two-flavor case (Fig.~\ref{fig:strong 2}), the oscillation probability $P_{eu}$ exhibits a smaller period compared to the weak-field results shown in Figs.~\ref{fig:2f_pi6_1m}. The presence of a nonzero Hu-Sawicki parameter $\lambda$ (yellow curve) induces noticeable phase shifts
 which become increasingly pronounced as the angular position  increases. For the three-flavor case (Fig.~\ref{fig:strong 3}), the oscillation period is also  smaller than that in the weak-filed regime. The oscillation amplitudes with zero and nonzero Hu-Sawicki parameter could be discriminated in this case.
	Our theoretical analysis highlights a potential interdisciplinary connection between neutrino oscillation physics and strong-field gravity; however, significant experimental and phenomenological work (including wave-packet decoherence, matter effects, and detector resolution) is required before such measurements become realistic.The enhanced sensitivity in high-curvature environments suggests that astrophysical neutrinos from compact objects could serve as valuable probes for testing modified gravity theories and for discriminating neutrino mass hierarchies.

	\section{Conclusion}\label{sec:conclusion}

	We have systematically investigated the effects of gravitational lensing on neutrino flavor oscillations within the spacetime geometry of  the Hu-Sawicki $f(R)$ gravity model. By deriving the covariant oscillation phase for both radial and lensed non-radial neutrino trajectories, we calculated the corresponding flavor transition probabilities for two and three-flavor neutrino oscillation probabilities under weak-field and strong-field regimes.

	The lensing-affected oscillation probabilities exhibit a distinct and quantifiable dependence on the three factors: the Hu-Sawicki model parameter
$\lambda$, the neutrino mass ordering (normal vs. inverted hierarchy), and the absolute value of the lightest neutrino mass
$m_l$. In the weak-field regime, the introduction of a nonzero $\lambda$ modulates the oscillation amplitude, while the mass ordering significantly influences both the magnitude and period of the probability curves. The value of $m_l$ further alters the oscillation profile, particularly affecting the period.
In the strong-field regime, the lensing effects become significant, with smaller oscillation periods and identifiable oscillation amplitudes to discriminate the gravity models.
We note that neutrinos are produced and detected as wave packets of finite width. Decoherence due to wave packet separation may affect oscillation probabilities, especially in curved spacetimes. In this work we adopted the plane-wave approximation to focus on gravitational lensing modifications. Decoherence effects in Schwarzschild spacetime have been studied in \cite{Alloqulov:2024sns}. Since our metric reduces to Schwarzschild when $\lambda=0$ and the deviation is small, the decoherence behavior is expected to be similar. A detailed analysis of wave-packet decoherence in the Hu-Sawicki $f(R)$ background is left for future work.

In summary, this work suggests that observations of lensed neutrino from compact astrophysical objects provide a  complementary avenue for probing fundamental neutrino properties such as the mass hierarchy and absolute mass scale and  for testing deviations from general relativity as encapsulated in modified gravity models like the Hu-Sawicki $f(R)$ gravity. The interplay between neutrino oscillation and strong-field gravity highlighted here shows a promising interdisciplinary frontier for the next generation of neutrino telescopes and multi-messenger astronomy.

For realistic astrophysical lenses such as the Sun, matter effects (the Mikheyev–Smirnov–Wolfenstein effect) can be significant and may dominate over gravitational lensing in some regimes. Our work focuses on pure vacuum gravitational lensing; a combined treatment including matter effects is an important direction for future research.Moreover, the finite energy and angular resolution of detectors can cause additional decoherence that may wash out the oscillatory patterns predicted here. Realistic assessments of observability must incorporate detector response functions. We leave such quantitative analyses to future work.To assess the observability, we make a rough estimate. Consider a typical blazar neutrino flux of $E^2 dN/dE \sim 10^{-12}\,\text{TeV}\,\text{cm}^{-2}\,\text{s}^{-1}$ at 10 TeV, and IceCube's effective area $\sim 1\,\text{m}^2$ at that energy. For a point source behind a compact lens, the lensed event rate is of order $10^{-3}$ per year, and the required angular resolution is $\sim 10^{-4}$ rad. These numbers indicate that with current detectors, detection is not feasible. However, with future large-area arrays and improved reconstruction, such measurements might become possible.

	\acknowledgments
	Ya-Ru Wang is grateful to Yu-Xuan Shi for his helpful discussions. Shu-Jun Rong thank X. Y. Chew, D. Stojkovic, K. S. Virbhadra, N. Tsukamoto, Tao Zhu and Chao Zhang, for communications on important works on gravity.
This work is supported by the National Natural Science Foundation of China under grant No.12065007.

	\bibliography{biblio}
\appendix

\section{Validity of the weak-field approximation used in this work}

In this appendix we clarify the conditions under which the weak-field expansions are justified. We emphasize that our weak-field approximation relies on the smallness of a different set of parameters, which are satisfied for all numerical examples presented in the main text.

\subsection{Small parameters in the expansion}

The following quantities are treated as small parameters throughout our derivation:

\begin{enumerate}
	\item $\displaystyle \varepsilon_1 = \frac{M}{r}$, where $r$ is the radial coordinate along the neutrino trajectory. This is the standard post-Newtonian parameter. For a solar-mass black hole and for $r \gtrsim r_0$ (the closest approach), we have $\varepsilon_1 \ll 1$ because $r_0$ is typically much larger than $M$ in the lensing configurations we consider.
	\item $\displaystyle \varepsilon_2 = \frac{b}{r_S}$ and $\displaystyle \varepsilon_3 = \frac{b}{r_D}$, where $b$ is the impact parameter and $r_S$, $r_D$ are the source and detector distances. In typical gravitational lensing, $b$ is of the same order as the Einstein radius, which is much smaller than $r_S$ and $r_D$. We have plotted the impact parameter \(b\) as a function of the source distance \(r_S\) and the detector distance \(r_D\), thereby demonstrating that our results are consistent with the weak-field approximation. The verification results for other cases are similar to those presented here.
	
	\begin{figure}[hb]
		\centering
		\includegraphics{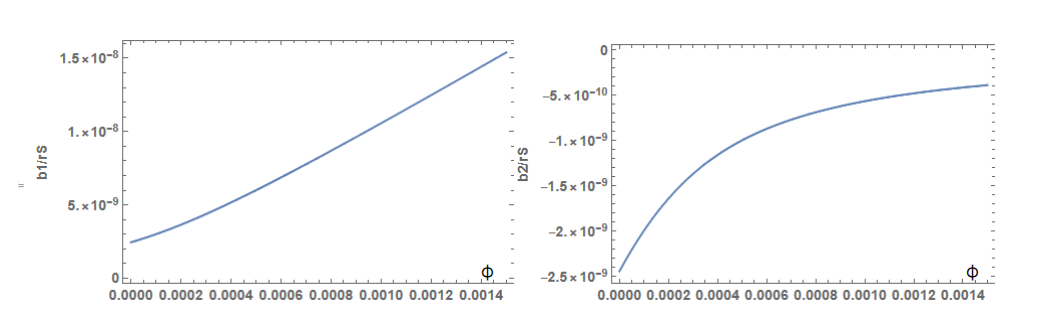}
		\caption{Three-flavor weak-field verification results with inverted mass ordering.}
		\label{fig:yan}
	\end{figure}
	
	\item $\displaystyle \varepsilon_4 = \frac{2M}{r_0}\,(1+\lambda r_0^2)$. This combination appears naturally in the relation between $b$ and $r_0$ (see Eq.~(32) and its derivation). Its smallness is guaranteed by $\varepsilon_1 = M/r_0 \ll 1$, irrespective of the magnitude of $\lambda r_0^2$. Even if $\lambda r_0^2$ is as large as $10^2$, the product $(M/r_0)\lambda r_0^2 = M\lambda r_0$ remains extremely small for astrophysical values of $M$ and $\lambda$. The chosen parameters in our paper satisfy the weak-field approximation conditions. Thus $\varepsilon_4$ is indeed a small parameter.
\end{enumerate}

\subsection{Why $\lambda r^2$ does not need to be small}

The parameter $\lambda r^2$ appears in the metric component $A(r)=1-2M/r+\lambda r^2$. In our weak-field expansions we \emph{do not expand} in powers of $\lambda r^2$. Instead, terms containing $\lambda r^2$ are kept in their exact form whenever they appear in denominators or under square roots. For instance, the exact phase integral (24) and the exact relation (31) contain $\lambda$ without any expansion. The expansions performed in Eqs.~(29) and (36) are only with respect to $M/r$ and $b/r$, while $\lambda$ remains as a fixed coefficient. Consequently, even when $\lambda r^2$ is not small, the expanded expressions remain valid as asymptotic series in $M/r$ and $b/r$, provided the small parameters listed above are indeed small. The numerical examples in the main text (including those where $\lambda r^2 \sim 100$) satisfy this condition.

\subsection{Conclusion of the appendix}

Thus, the weak-field approximation used in this paper is fully justified under the conditions $\varepsilon_1,\varepsilon_2,\varepsilon_3,\varepsilon_4 \ll 1$, which hold for all our numerical examples. The fact that $\lambda r^2$ may be large does not affect the consistency of the expansion, because $\lambda$ is never treated as a small expansion parameter.

\section{Simplified derivation of the weak-field deflection angle}

In this appendix we present a detailed simplification of the deflection angle expression given in Eq.~(23) of Ref.~[55], leading to the compact formula used in the main text, Eq.~(36).

\subsection{Parameter definitions}

The Hu-Sawicki $f(R)$ black hole parameter $\lambda$ is defined as
\[
\lambda = \frac{m^2}{12} \left( \frac{n-2}{2c_2} \right)^{\frac{1}{n}}.
\]
To simplify the expansion coefficients we introduce the auxiliary quantity
\[
A \equiv m^2 \left( \frac{n-2}{c_2} \right)^{\frac{1}{n}}.
\]
From the definitions one finds $A = 12 \cdot 2^{\frac{1}{n}} \lambda$. Consequently, any combination of the form $2^{-k-\frac{1}{n}} A$ reduces to
\[
2^{-k-\frac{1}{n}} A = 12 \cdot 2^{-k} \lambda. \tag{B1}
\]

\subsection{Taylor expansions in the far-distance limit}

In the weak-field far-distance limit, the source and the receiver are located at distances much larger than the impact parameter, i.e. $u_R = 1/r_R \to 0$ and $u_S = 1/r_S \to 0$. Let the impact parameter be $b$ and denote $\epsilon = 1/b$ (note that in the final result $\epsilon$ will be replaced by $1/b$). We employ the following Taylor expansions (set $\delta = \epsilon u$):
\[
\begin{aligned}
	\arcsin \delta &= \delta + \frac{\delta^3}{6} + O(\delta^5),\\[2mm]
	\frac{1}{\sqrt{1-\delta^2}} &= 1 + \frac{1}{2}\delta^2 + \frac{3}{8}\delta^4 + O(\delta^6),\\[2mm]
	\frac{1}{(1-\delta^2)^{3/2}} &= 1 + \frac{3}{2}\delta^2 + \frac{15}{8}\delta^4 + O(\delta^6),\\[2mm]
	\frac{1}{(1-\delta^2)^{5/2}} &= 1 + \frac{5}{2}\delta^2 + \frac{35}{8}\delta^4 + O(\delta^6).
\end{aligned}
\]

\subsection{Term-by-term simplification}

The deflection angle in Eq.~(23) of Ref.~[55] can be written as a sum of terms $\hat{\alpha} = T_1 + T_2 + \cdots + T_9$. We simplify each term below.

\paragraph{Term $T_1$:}
\[
T_1 = \left[\frac{15M^2}{4\epsilon} + 5 \cdot 2^{-4-\frac{1}{n}} M^2 A\right] \left(\pi - \arcsin(\epsilon u_R) - \arcsin(\epsilon u_S)\right).
\]
Using (B1), $2^{-4-1/n}A = 12 \cdot 2^{-4} \lambda = \frac{3}{4}\lambda$. For $u_R,u_S\to 0$,
\[
\pi - \arcsin(\epsilon u_R) - \arcsin(\epsilon u_S) = \pi - \epsilon(u_R+u_S) + O(u^3).
\]
Hence,
\[
T_1 = \pi\left(\frac{15M^2}{4b^2} + \frac{15}{4}\lambda M^2\right) + \text{finite-distance corrections}.
\]

\paragraph{Terms $T_2$, $T_3$, $T_4$:}
These terms contain higher powers of $u_R$ or $u_S$ and vanish in the far-distance limit: $T_2, T_3, T_4 \to 0$.

\paragraph{Term $T_5$:}
\[
T_5 = \left[\frac{2M}{\epsilon} - 2^{-3-\frac{1}{n}} \epsilon M A\right] \left(\frac{1}{(1-\epsilon^2 u_R^2)^{3/2}} + \frac{1}{(1-\epsilon^2 u_S^2)^{3/2}}\right).
\]
Using (B1), $2^{-3-1/n}A = 12 \cdot 2^{-3} \lambda = \frac{3}{2}\lambda$. At $u=0$, $\frac{1}{(1-\epsilon^2 u^2)^{3/2}} = 1$, thus
\[
T_5 = 2\left(\frac{2M}{b} - \frac{3}{2}\lambda b M\right) = \frac{4M}{b} - 3\lambda b M.
\]

\paragraph{Terms $T_6$, $T_7$, $T_8$:}
These terms vanish as $u\to 0$.

\paragraph{Term $T_9$:}
\[
T_9 = -\frac{2^{-3-\frac{1}{n}}}{3} \epsilon A \left(\frac{u_R^{-1}}{\sqrt{1-\epsilon^2 u_R^2}} + \frac{u_S^{-1}}{\sqrt{1-\epsilon^2 u_S^2}}\right).
\]
Using the expansion
\[
\frac{u^{-1}}{\sqrt{1-\epsilon^2 u^2}} = u^{-1} + \frac{1}{2}\epsilon^2 u + O(u^3),
\]
the dominant part as $u\to 0$ is $u^{-1}$, which diverges. In a realistic astrophysical configuration (e.g., the Sun–Earth system) $u_R$ and $u_S$ are small but nonzero. This term contributes a constant offset independent of $b$ (since $\epsilon A \propto \lambda b$, while $1/u$ is $b$-independent). Therefore it does not affect the relative variation of the oscillation probability with $b$, and we omit it in the leading-order expression.

\subsection{Final deflection angle}

Collecting all the non-vanishing contributions and neglecting higher-order terms, we obtain the weak-field deflection angle
\[
\hat{\alpha} = \frac{15\pi M^2}{4b^2} + \frac{15\pi \lambda M^2}{4} + \frac{4M}{b} - 3\lambda b M + \text{finite-distance corrections}.
\]
This is precisely the expression used in Eq.~(36) of the main text. Thus, starting from the cumbersome expression in Ref.~[55] and applying Taylor expansions together with the parameter simplifications above, we have derived the compact deflection angle formula employed in our numerical analysis.

\end{document}